\documentclass[aip,apl,twocolumn,amsmath,amssymb,reprint]{revtex4-1}

\usepackage{graphicx}
\usepackage{dcolumn}
\usepackage{bm}

\preprint{v6.8}

\begin{document}
\title{AlGaAs/GaAs single electron transistors fabricated without modulation doping}

\author{A.M. See}
\author{O. Klochan}
\author{A.R. Hamilton}
\email{Alex.Hamilton@unsw.edu.au}
\author{A.P. Micolich}
\affiliation{School of Physics, University of New South Wales,
Sydney NSW 2052, Australia}

\author{M. Aagesen}
\author{P.E. Lindelof}
\affiliation{Nanoscience center, University of Copenhagen,
Universitetsparken 5, DK-2100 Copenhagen, Denmark}

\date{\today}

\begin{abstract}
We have fabricated quantum dot single electron transistors, based on
AlGaAs/GaAs heterojunctions without modulation doping, which exhibit
clear and stable Coulomb blockade oscillations. The temperature
dependence of the Coulomb blockade peak lineshape is well described
by standard Coulomb blockade theory in the quantum regime. Bias
spectroscopy measurements have allowed us to directly extract the
charging energy, and showed clear evidence of excited state
transport, confirming that individual quantum states in the dot can
be resolved.
\end{abstract}
\maketitle

Quantum dots are central to nanoelectronics and have been used to
realize single-electron transistors (SETs),~\cite{Meirav:1990prl}
artificial atoms,~\cite{Kastner:1993phystod} ultra-sensitive
electrometers,~\cite{Fujisawa:2004apl} and may ultimately be used as
elements for quantum information applications.~\cite{Loss:1998pra,
Hanson:2007rmp} Semiconductor quantum dots are typically defined
using negatively-biased surface gates to deplete regions of the
two-dimensional electron gas (2DEG) formed in a modulation-doped
AlGaAs/GaAs heterostructure.~\cite{Kouwenhoven:1997nato} Although
modulation doping results in high electron
mobilities,~\cite{Dingle:1978apl} it can also cause significant
charge noise and temporal instability due to rapid switching of the
dopants between ionized and de-ionized states.~\cite{Kurdak:1997prb,
Pioro-Ladriere:2005prb, Buizert:2008prl} Methods such as cooling the
device with the gates biased~\cite{Pioro-Ladriere:2005prb} or
depositing gates on a thin insulating layer~\cite{Buizert:2008prl}
can reduce charge noise but not eliminate it entirely, hindering the
development of ultra-sensitive quantum devices.

Here we report the development of a quantum dot in a heterostructure
without modulation doping. Instead the electron gas is `induced'
electrostatically using a degenerately doped metallic top-gate in an
otherwise undoped heterostructure.~\cite{Kane:1995apl} The present
device architecture overcomes many of the limitations of the
inverted semiconductor-insulator-semiconductor (ISIS)
heterostructure used in some of the earliest studies of GaAs
SETs.~\cite{Meirav:1990prl} In the ISIS device, a heavily-doped
substrate was used as a gate to induce a 2DEG at an inverted
AlGaAs/GaAs interface, with the quantum dot defined by
negatively-biased Schottky gates. However, the ISIS heterostructure
also has a delta-doped layer between the Schottky gates and the 2DEG
to counter GaAs surface states, leading to similar disorder to
modulation-doped heterostructures.~\cite{Meirav:1990thesis} ISIS
devices also suffer from the reduced mobility inherent in inverted
interfaces,~\cite{Heiblum:1985jvstb} and ohmic contact penetration
into the doped substrate.~\cite{Meirav:1990thesis} Our device
overcomes these problems -- there is no need for shallow ohmic
contacts, no inverted interface, and although we use a doped cap
layer as a gate, it is degenerately doped to have a metallic
conductivity at low temperature, providing a sufficiently high
electron density to screen the 2DEG from ionized donors in the cap.
With this device we observe clean and stable Coulomb blockade
oscillations, and single particle states within the dot.

Our devices were fabricated from an AlGaAs/GaAs heterostructure
consisting of (from undoped GaAs buffer upwards): a $160$~nm undoped
AlGaAs barrier, a $25$~nm GaAs spacer, and a $35$~nm n$^{+}$-GaAs
cap used as a metallic gate. Hall bars with annealed NiGeAu ohmic
contacts are defined using a self-aligned
process.~\cite{Kane:1995apl} The 2DEG is depleted at top gate bias
$V_{TG} < 0.32$~V, and above this threshold, the electron density $n
= (-1.09 + 3.41 V_{TG}) \times 10^{11}$~cm$^{-2}$. Characterization
of the heterostructure gave a mobility of $\sim 300,000$~cm$^{2}/$Vs
at $n \sim 1.8 \times 10^{11}$~cm$^{-2}$. A quantum dot with
dimensions $0.54 \times 0.47~\mu$m was defined by using electron
beam lithography and a H$_{2}$SO$_{4}$ etch to form a $\sim 45$~nm
deep trench dividing the cap into seven separate gates, as shown in
Fig.~1(c). The quantum dot and adjacent 2DEG reservoirs are
populated using the top gate. At each end is a quantum point contact
(QPC) defined by split-gates biased at $V_{L}$ and $V_{R}$, which
control the tunnel barriers between the dot and the reservoirs (both
QPCs are set with a conductance less than $2e^{2}/h$). Finally, the
bottom `plunger gate', biased at $V_{PG}$, allows the occupancy of
the dot to be tuned (the upper plunger gate was connected to ground
for the entire experiment). Electrical measurements were performed
using ac and dc techniques with the dot cooled to millikelvin
temperatures by a dilution refrigerator. The temperature $T$ was
measured using a Nanoway cryoelectronics primary Coulomb blockade
thermometer mounted with the device.

\begin{figure}[h]
\includegraphics[width=8cm]{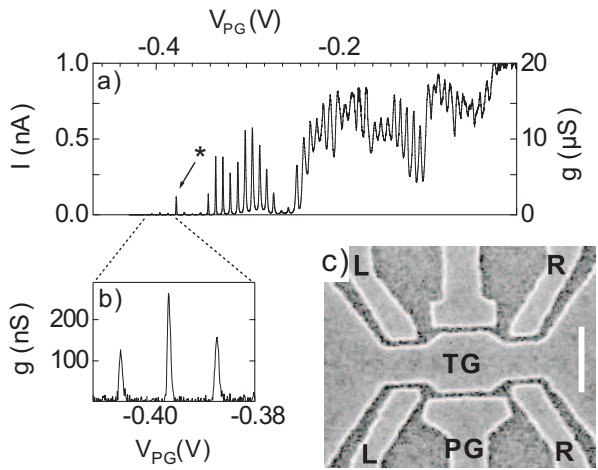}
\caption{Current $I$ (left axis) and two-terminal conductance $g$
(right axis) vs plunger gate voltage $V_{PG}$ at $T = 40$~mK
measured with an ac excitation of $V_{ac} = 50~\mu$V. The *
indicates the Coulomb Blockade (CB) peak presented in Fig.~4 (b) A
close-up view of the last few resolvable CB oscillations. (c) An
electron-beam micrograph of the device. The letters indicate the
various gates referred to in the text. The scale bar is $500$~nm in
length.}
\end{figure}

Figure~1(a) shows CB oscillations as the plunger gate is used to
tune the dot occupancy. Similar oscillations are obtained by
sweeping the other gates. Estimates of the respective gate
capacitances can obtained from an analysis of the periodicity these
oscillations (ignoring the discrete energy level structure of the
dot). The capacitances obtained are $17.0$~aF, $14.5$~aF, $20$~aF
and $107$~aF for the left QPC, right QPC, plunger gate and top gate
respectively. The top gate is the dominant contribution at $67 \%$
of the total gate capacitance of $160$~aF. We estimate that the dot
contains at most $300$ electrons, based on the dot geometry,
electron density and a $50$~nm depletion region at the dot walls,
giving an energy level spacing $\Delta E \approx 45~\mu$eV. The CB
oscillations in this device are quite sharp, and as shown in the
close-up view of the last few resolvable CB peaks in Fig. 1(b), the
conductance falls to zero for extended stretches between the CB
peaks.

Figure 2 shows a color-map of the conductance $g$ versus right QPC
gate bias $V_{R}$ and plunger gate bias $V_{PG}$. Bright regions
mark the CB peaks, and the current is blockaded in the dark regions.
The data in Fig.~1(a) corresponds to a slice along the horizontal
white dashed line in Fig.~2. There is no evidence of charge trapping
or random telegraph noise, which would produce discontinuous jumps
in the bright lines in Fig.~2. In an ideal device free of crosstalk
between the gates, the bright lines would be vertical; their
relatively large slope indicates strong crosstalk between the right
QPC and the plunger gate (similar data is obtained for the left
QPC). This crosstalk is unavoidable due to the close proximity and
similar capacitances of the QPC and plunger gate, but might be
reduced with further optimization of the device design.

\begin{figure}
\includegraphics[width=8cm]{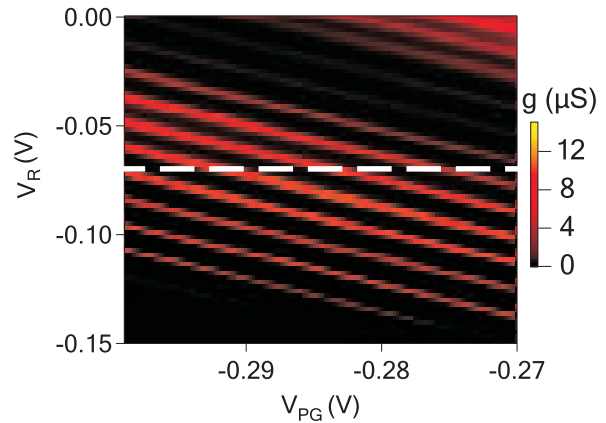}
\caption{A color-map of the conductance $g$ versus the right QPC
bias $V_{R}$ and plunger gate bias $V_{PG}$, obtained with $V_{TG} =
0.85$~V and $V_{L} = 0$~V. The data in Fig. 1(a) corresponds to a
slice along the horizontal white dashed line.}
\end{figure}

Bias spectroscopy measurements are shown in Fig.~3, where the
conductance is plotted as a color-map versus the dc source-drain
bias $V_{SD}$ and $V_{PG}$. Dark regions indicate low $g$, and form
a sequence of `Coulomb diamonds' (highlighted by the white solid
lines) where current through the dot is
blockaded.~\cite{Kouwenhoven:2001rpp} Again, there is no evidence of
charge noise or random switching events in Fig.~3, demonstrating the
stability of our device. The bright regions outside the Coulomb
diamonds running parallel to the diamond edge (highlighted by
white/black dashed lines) suggest transport via excited states in
the dot.~\cite{DeFrancheschi:2001prl} The level spacing $\Delta E$
between the ground and excited states can be measured by the
separation in $V_{SD}$ between the excited state line and the
diamond edge, with $\Delta E \sim 180$ to $240~\mu$eV. The level
spacing is larger than the $45~\mu$eV estimate obtained earlier,
suggesting that the dot contains significantly fewer than $300$
electrons. This is likely due to lateral depletion caused by the
shallow etch and fringing fields from the plunger and QPC gates. The
charging energy can be directly extracted from an analysis of the
Coulomb diamonds by subtracting $\Delta E$ from the half-diamond
height,~\cite{Hanson:2007rmp} and ranges between $0.44$ and
$0.45$~meV, varying slightly with dot occupancy. This charging
energy corresponds to a total dot capacitance of $\sim 360$~aF,
approximately consistent with the sum of the gate capacitances.

\begin{figure}
\includegraphics[width=8cm]{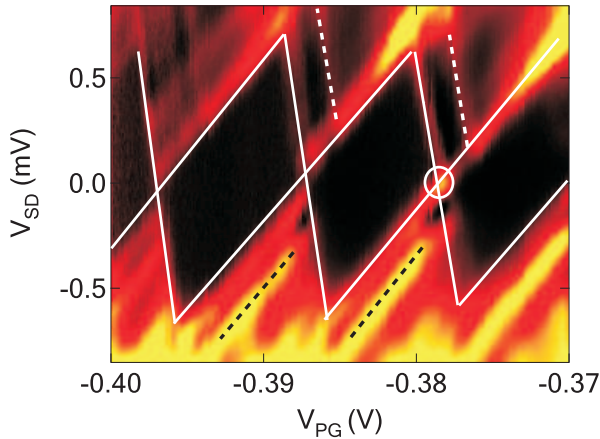}
\caption{Bias spectroscopy of the quantum dot, showing the
conductance $g$ (color axis) against plunger gate voltage $V_{PG}$
($x$-axis) and dc source-drain bias $V_{SD}$ ($y$-axis). The dark
regions correspond to $g = 0$, and form `Coulomb diamonds'
(highlighted by solid white lines). Dashed lines indicate regions
where transport via excited states occurs. The CB peak highlighted
by the * in Fig.~1 and analyzed in Fig.~4 is located inside the
white circle.}
\end{figure}

Further information was obtained from the temperature dependence of
the CB peaks, in particular, from an analysis of their lineshape it
is possible to differentiate whether the dot is in the classical or
quantum transport regime.~\cite{Beenakker:1991prb,Folk:1996prl} In
the classical regime $\Delta E \ll k_{B}T \ll E_{C}$, the peak
conductance $g_{peak}(T)$ is temperature independent and the peak
full width at half maximum $w$ increases linearly with $T$. In the
quantum regime $\Delta E > k_{B}T$, $g_{peak}(T)$ goes as $1/T$
instead, with the linear relationship between $w$ and $T$
maintained.~\cite{Beenakker:1991prb} More rigorously:

\begin{equation}
\frac{g}{g_{peak}} \approx cosh^{-2}(\frac{\alpha e \Delta V}{2k_{B}T})
\end{equation}

\noindent in the quantum regime, where $\alpha = (C_{PG}/\Sigma
C_{g})$ is the plunger gate `lever arm' and $\Delta V =
|V_{PG}-V_{PG}^{peak}|$ is the plunger gate voltage relative to the
center of the CB peak.

\begin{figure}
\includegraphics[width=8cm]{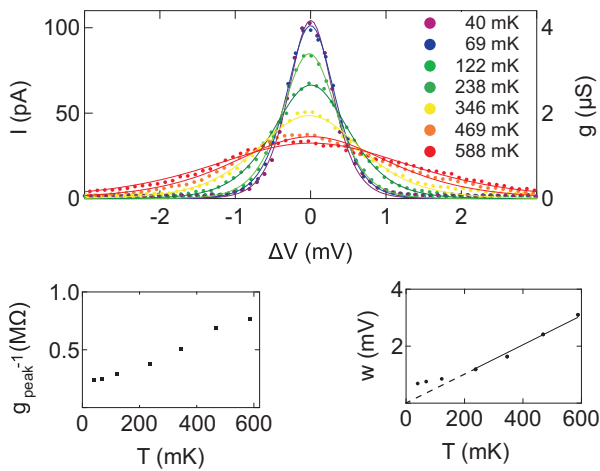}
\caption{(a) Temperature dependence of the CB peak centered at
$V_{PG} \approx -0.38$~V and indicated by the * in Fig.~1(b). The
current $I$ (left axis) and conductance $g$ (right axis) are plotted
against the voltage $\Delta V$ relative to the peak's center. The
solid lines are fits of Eqn.~1 to the experimental data (points).
(b) The inverse peak conductance $g^{-1}_{peak}$ versus $T$. The
linear trend demonstrates that the dot is in the quantum regime. (c)
The peak full width at half maximum $w$ versus $T$.}
\end{figure}

In Fig.~4(a), we show the temperature dependence for the CB peak
centered at $V_{PG} \approx -0.38$~V for temperatures between $40$
and $590$~mK. The solid lines in Fig.~4(a) are fits of Eqn.~1 to the
data with $\alpha$ and $g_{peak}$ as free parameters. In Fig.~4(b)
we plot $g^{-1}_{peak}$ versus $T$, with the linear trend confirming
that our dot is in the quantum regime. In Fig.~4(c), we plot $w$
against $T$, and from the gradient $dw/dT$ of the linear trend at
higher $T$ we can extract the lever arm using the relationship
$\alpha = 4k_{B}ln(\sqrt{2}+1)/e \times dT/dw$.~\cite{Folk:1996prl}
We obtain $\alpha = 0.059$, which is consistent with $0.066$, the
average of the lever arms extracted from the left and the right
diamonds of the same CB peak (Fig. 3). Finally, in Figs.~4(b/c) we
observe saturation of both $g^{-1}_{peak}$ and $w$ as $T \rightarrow
0$, providing an estimate of the minimum electron temperature in our
device of $\sim 140$~mK.

In summary, we have fabricated quantum dots in AlGaAs/GaAs
heterostructures without modulation doping. We use a heavily-doped
cap layer, patterned into gates by electron-beam lithography and wet
etching, to electrostatically control the electron population of the
dot. Our device shows clear, stable CB oscillations in the quantum
regime with transport via excited states in the dot also apparent.
The improved noise performance afforded by removing the modulation
doping makes this device architecture interesting for applications
such as quantum information and ultra-sensitive electrometry where
very stable quantum dots are extremely useful.

We thank M.A. Eriksson for helpful discussions. We acknowledge
funding from the Australian Research Council (DP0772946). ARH
acknowledges an ARC Professorial Fellowship.

\end{document}